\documentclass{jfm}
\usepackage{graphicx}
\usepackage{epstopdf, epsfig}
\usepackage{float}
\usepackage{amsfonts}
\usepackage{bm}
\usepackage{hyperref}
\usepackage{mathtools}
\usepackage{color}

\usepackage[utf8]{inputenc}
\usepackage[english]{babel}

\usepackage[dvipsnames]{xcolor}
\colorlet{LightRubineRed}{RubineRed!70!}
\colorlet{Mycolor1}{green!10!orange!90!}
\definecolor{Mycolor2}{HTML}{00F9DE}

\shorttitle{Entropy and fluctuation relations in  
turbulence}
\shortauthor{H. Yao, T.A. Zaki and C. Meneveau}

\title{Entropy and fluctuation relations in 
isotropic
turbulence}

\author{H. Yao\aff{1},
  T.A. Zaki\aff{1} \and 
  C. Meneveau\aff{1}}

\affiliation{\aff{1}Department of Mechanical Engineering \& IDIES, Johns Hopkins University}

\begin{document}

\maketitle

\begin{abstract}
Based on a generalized local Kolmogorov-Hill equation expressing the evolution of kinetic energy integrated over spheres of size $\ell$ in the inertial range of fluid turbulence, we examine a possible definition of entropy and entropy generation for turbulence. Its measurement from direct numerical simulations in isotropic turbulence leads to confirmation of the validity of the fluctuation relation (FR) from non-equilibrium thermodynamics in the inertial range of turbulent flows. Specifically,  the ratio of probability densities of forward and inverse cascade at scale $\ell$ is shown to follow exponential behavior with the entropy generation rate if the latter is defined by including an appropriately defined notion of ``temperature of turbulence'' proportional to the kinetic energy at scale $\ell$.    
\end{abstract}

\maketitle
 \section{Introduction}
A long-standing hope of research in turbulence is that connections to non-equilibrium thermodynamics and statistical mechanics could be established.  For example, connections were attempted some time ago for vortex filament models \citep{chorin1991equilibrium}, infinitely divisible cascade processes (see \citep{castaing1996temperature} and references therein), as well asmultifractal models of the energy cascade with its analogues to Gibbs free energy, Legendre transformations \citep{paladin1987anomalous,chhabra1989extraction}, and even phase transitions
\citep{meneveau1990two}. However, connections between such  models of the cascade   and the Navier-Stokes equations remain tenuous to this day. More recently, considering the reversibility of Navier-Stokes equations in the inviscid limit (or in the inertial range of turbulence) and building upon prior works by \cite{she1993constrained,carati2001modelling,cichowlas2005effective,domaradzki2007analysis,eyink2009localness,cardesa2015temporal,cardesa2017turbulent}, an analysis of the cascade process and possible connections to entropy was carried out by \cite{vela2021entropy}. Various consequences of the time-reversibility of the inertial range dynamics were explored and connections were made to physical-space flow structures in seeking physical explanations for the asymmetry between positive (forward) and negative (inverse) cascade rates.    
Recently \cite{fuchs2020small} proposed a definition of entropy change of individual cascade trajectories  based on a Fokker-Planck stochastic model equation and tested predictions from non-equilibrium thermodynamics. Similarly, \cite{porporato2020fluctuation} considered fluctuations in spectral models in Fourier space. We here explore a new definition of entropy generation rate 
based on the exact kinetic energy transport equation in the inertial range of turbulence and  test quantitative predictions from non-equilibrium thermodynamics regarding the direction and magnitude of the cascade rate. 

 \section{The generalized Kolmogorov-Hill equation for local kinetic energy}
The kinetic energy of turbulence can be defined using structure functions \citep{frisch1995turbulence}. As a generalization of the celebrated Karman-Howarth and Kolmogorov equations for structure functions, \cite{hill2001equations, hill2002exact} derived what will here be denoted as the generalized Kolmogorov-Hill equation (GKHE).  It is obtained from the incompressible Navier-Stokes equations written at two points and before averaging, it accounts for the local time evolution of velocity increment magnitude (square) at a specific physical location ${\bf x}$ and  scale ${\bf r}$, and incorporates effects of viscous dissipation, viscous transport, advection, and pressure \citep{hill2001equations, hill2002exact}. 
With no mean flow and for scales at which large-scale forcing can be neglected, the instantaneous GKHE reads  
\begin{equation}
\frac{\partial \delta u _i^2}{\partial t} + u^*_{j}\frac{\partial \delta u _i^2}{\partial x_j}  = 
-\frac{\partial \delta u _j\delta u _i^2}{\partial r_j}-\frac{8}{\rho}\frac{\partial p^*\delta u _i}{\partial r_i} 
+\nu \frac{1}{2} \frac{\partial^2 \delta u _i  \delta u _i}{\partial x_j \partial x_j
}+ 
2\nu \frac{\partial^2 \delta u _i \delta u _i}{\partial r_j \partial r_j}
-
4\epsilon^*,
\label{ins_KHMH_noint}
\end{equation}
where $\delta u_i = \delta u_i({\bf x};{\bf r}) =  u_i^+ - u_i^-$ is the velocity increment vector in the ith Cartesian direction. The superscripts $+$ and $-$ represent two points ${\bf x}+{\bf r}/2$ and ${\bf x}-{\bf r}/2$ in the physical domain that have a separation vector $r_i = x^+_i - x^-_i$ and middle point  
$x_i = (x^+_i + x^-_i)/2$. The superscript $*$ denotes the average value between two points. For instance,  the two-point average dissipation is defined as $\epsilon^* = (\epsilon^+ +\epsilon^-)/2$. Here $\epsilon^{\pm}$ is the ``pseudo-dissipation'' defined locally as $\epsilon =\nu  ({\partial u_i}/{\partial x_j})^2$, where $\nu$ is the kinematic fluid viscosity. 

As already noted by  \cite{hill2002exact} (\S 3.5), Eq. \ref{ins_KHMH_noint} at any point ${\bf x}$ can be integrated  over a sphere in {\bf r}-space, up to a diameter which here will be denoted as the scale $\ell$ 
(it will be assumed to be in the inertial range so that viscous diffusion terms are neglected \cite{yao2023comparing}). The resulting equation is divided by the volume of the sphere $(V_\ell=\frac{4}{3}\pi( {\ell}/{2})^3)$ and a factor of 4, which yields its integrated form,
\begin{equation}
  \frac{\widehat{d}k_\ell}{d t}   =  \Phi_\ell  +   P_\ell  - \epsilon_\ell,
 \label{KHMH_local}
\end{equation}
where 
\begin{equation}
 \frac{ {\widehat{d}} k_\ell}{d t} \equiv 
 \frac{1}{2 \,V_r}\int\limits_{V_{r}} \left(\frac{\partial \frac{1}{2}\delta u _i^2}{\partial t} + u^*_{j} \frac{\partial \frac{1}{2}\delta u _i^2}{\partial x_j} \right) \, d^3{\bf r}_s =  \frac{ \partial k_\ell}{\partial t}+
 \frac{1}{2 \,V_r}\int\limits_{V_{r}}  u^*_{j} \frac{\partial \frac{1}{2}\delta u _i^2}{\partial x_j}  \, d^3{\bf r}_s,
 \label{timederiv1}
\end{equation}
is a local time rate of change of kinetic energy at all scales smaller or equal to $\ell$. We have defined the kinetic energy associated to the scales smaller than $\ell$ according to
\begin{equation}
 k_\ell({\bf x},t) \equiv 
 \frac{1}{2 \,V_\ell}\int\limits_{V_{\ell}}  \frac{1}{2} \delta u_i^2({\bf x},{\bf r}) \, d^3{\bf r}_s,
 \label{kinetic}
\end{equation}
where the 1/2 factor in front of the integral accounts for the fact that a volume integration over the sphere $V_\ell$ of diameter $\ell$ will  count the increments $\delta u _i^2$ twice. The quantity $k_\ell({\bf x},t)$ will be central to our analysis.
Eq. \ref{KHMH_local} also includes
\begin{equation}
\epsilon_\ell({\bf x}) \equiv \frac{1}{V_\ell}\int\limits_{V_{\ell}}
 \epsilon^*({\bf x},{\bf r}) d^3{\bf r}_s,
\end{equation}
the locally volume averaged rate of dissipation envisioned in the the \cite{kolmogorov1962refinement} refined similarity hypothesis (KRSH). The radius vector ${\bf r}_s = {\bf r}/2$ is integrated up to magnitude $\ell/2$, and 
\begin{equation}
\Phi_\ell  \equiv -\frac{3}{4\,\ell}\frac{1}{S_\ell}\oint\limits_{S_{\ell}} \delta u _i^2\,\delta u _j\,  \hat{r}_j dS = -\frac{3}{4\,\ell} \, [\delta u_i^2 \delta u_j \hat{r}_j ]_{S_\ell} 
\end{equation}
is interpreted as the local energy cascade rate in the inertial range at scale 
$\ell$ at position ${\bf x}$. Note that Gauss theorem is used to integrate the first term on the RHS of Eq. \ref{ins_KHMH_noint} over the $r_s$-sphere's surface, with area element $\hat {r}_j dS$, with $\hat {\bf r} = {\bf r}/|{\bf r}|$, and $S_\ell = 4\pi (\ell/2)^2$ the sphere's overall area (care must be taken as the Gauss theorem applies to the sphere's radius vector ${\bf r}_s = {\bf r}/2$ and 
$\partial_{r} = 2 \, \partial_{r_s}$). Averaging over the surface $S_\ell$  is denoted by $[...]_{S_\ell}$. Finally, Eq. \ref{KHMH_local} also includes  
\begin{equation}
P_\ell  \equiv -\frac{6}{\ell}\frac{1}{S_\ell}\oint\limits_{S_{\ell}} \frac{1}{\rho} \, p^* \, \delta u _j \, \hat{r}_j \,dS,
\end{equation}
the surface averaged pressure work term at scale $\ell$ (defined as positive if the work is done {\it on} the system inside the volume $V_\ell$).  
Equation \ref{KHMH_local} is local (valid at any point ${\bf x}$ and time $t$), and each of the terms in the equation can be evaluated from data according to their definition using a sphere centered at any middle point ${\bf x}$.  For more details about this formulation, see \cite{yao2023comparing}.

In most prior works, it is the statistical average of Eq. \ref{KHMH_local} that is considered \citep{
Monin_Yaglom_1975,danaila_anselmet_zhou_antonia_2001,Danaila_et_al_2012,
carbone2020vortex}.
Using ensemble averaging for which isotropy of the velocity increment statistics can be invoked, in the inertial range neglecting the viscous term, the rate of change and pressure terms vanish and one recovers the Kolmogorov equation for two-point longitudinal velocity increments that connects third-order moments to the overall mean rate of viscous dissipation via the celebrated $-4/5$ law: $\langle \delta u_L^3(\ell) \rangle = - \frac{4}{5} \ell \langle \epsilon \rangle$ \citep{kolmogorov1941local,frisch1995turbulence}.
Here $\langle .. \rangle$ means global averaging, $\delta u_L(\ell)$ is the longitudinal  velocity increment over distance $\ell$, assumed to be well inside the inertial range of turbulence.  Without averaging, and also without the viscous, pressure and unsteady terms, Eq. \ref{KHMH_local} becomes the ``local 4/3-law'' obtained by  \cite{duchon2000inertial} and discussed by  \cite{eyink2002local} and \cite{dubrulle2019beyond}, connecting $\Phi_\ell$ to $\epsilon_\ell$ in the context of energy dissipation in the $\nu \to 0$ limit (see \cite{yao2023comparing} regarding subtle differences with Hills's more symmetric two-point approach used here).

Returning to the time derivative term in Eq. \ref{timederiv1}, in order to separate advection due to overall velocity at scale $\ell$ and smaller scale contributions, we define the filtered advection velocity as $ 
 \tilde{u}_j \equiv 
 \frac{1}{V_\ell}\int_{V_{\ell}}   u_j^* \, d^3{\bf r}_s$.
It corresponds to a filtered velocity using a spatial radial top-hat filter \citep{yao2023comparing}. Accordingly, we may write
\begin{equation}
 \frac{ {\widehat{d}}k_\ell}{d t}  =   \frac{ {\widetilde {d}} k_\ell}{d t} + \frac{\partial q_j}{\partial x_j} = \frac{ \partial k_\ell}{\partial t} +  \tilde{u}_j \, \frac{ \partial k_\ell }{\partial x_j} + \frac{\partial q_j}{\partial x_j},
 \label{timederiv2}
 \end{equation}
 where $\tilde{d}/dt = \partial/\partial t + \tilde{u}_j\partial/\partial x_j$ and 
$ q_j = 
 \frac{1}{V_\ell}\int_{V_{\ell}} \frac{1}{2} \left(\delta u _i^2 \delta u^*_j \right) \, d^3{\bf r}_s$ (the spatial flux of small-scale kinetic energy), with $\delta u^*_j \equiv u^*_j - \tilde{u}_j$. 
The evolution of kinetic energy of turbulence at scales at and smaller than $\ell$ (in the inertial range, i.e. neglecting viscous diffusion and forcing terms) is thus given by
\begin{equation}
  \frac{\widetilde{d} k_\ell}{d t}   =  \Phi_\ell  - \epsilon_\ell + P_\ell - \frac{\partial q_j}{\partial x_j},
 \label{KHMH_local2}
\end{equation}
This 
equation represents the ``first law of thermodynamics'' for our system of interest. The system can be considered to be the eddies inside the sphere of diameter $\ell$ consisting of turbulent fluid (see Fig. \ref{fig:sketchab}(a)). We consider the smaller-scale turbulent eddies inside the sphere to be analogous to a set of interacting ``particles'' which are exposed to energy exchange with the larger-scale flow structures at a rate $\Phi_\ell$, loosing energy to molecular degrees of freedom at a rate $\epsilon_\ell$, and also being exposed to work per unit time done by pressure at its periphery ($P_\ell$). Spatial turbulent transport (spatial flux $q_j$) can also be present.

\section{Analogy with Gibbs equation and definition of entropy}
The energetics (first law Eq. \ref{KHMH_local2}) of the system of eddies inside the ball of size $\ell$ invites us to write a sort of Gibbs equation, in analogy to the standard expression 
\begin{equation}
T ds = d{e} + p\,dv,
\label{eq:TDS}
\end{equation}
where $T$ is temperature, aiming to define an entropy $s$.  The internal energy ${e}$ is analogous to $k_\ell$ and the pressure work ($p\,dv$, work done {\it by} the system) is analagous to $-P_\ell$ since the volume change $dv$ is the surface integration of $\delta u_j \hat{r}_j$ times a time increment $dt$. Rewritten as a rate equation (i.e., dividing by $dt$),  the analog to Gibbs equation for our system reads
\begin{equation}
  T \,\frac{\widetilde{d} s_\ell}{d t}   =  \frac{\widetilde{d} k_\ell}{d t} - P_\ell,
 \label{KHMH_gibbs1}
\end{equation}
where $s_\ell$ is a new quantity defined via this equation and is akin to an entropy (intensive variable) of the system of small-scale eddies inside the sphere of diameter $\ell$. Also, $T$ has to be some suitably defined temperature. 
Combining Eq. \ref{KHMH_gibbs1} with the energy equation (Eq. \ref{KHMH_local2}) one obtains
\begin{equation}
   \frac{\widetilde{d} s_\ell}{d t}   = \frac{1}{T} \left(\Phi_\ell - \epsilon_\ell  - \frac{\partial q_j}{\partial x_j} \right).
 \label{KHMH_gibbs2}
\end{equation}
The heat exchange with the ``thermal reservoir'' (here considered to be the molecular degrees of freedom inside the sphere) at rate $\epsilon_\ell$ also occurring at temperature $T$ then generates a corresponding change (increase) of entropy of the ``reservoir'' at a rate 
\begin{equation}
  \frac{\widetilde{d} s_{\rm res}}{d t}   = \frac{ \epsilon_\ell}{T}.
 \label{entropybath}
\end{equation}
The generation rate of {\it total entropy} $s_{\rm tot} = s_{\ell}+s_{\rm res}$ is then given by
\begin{equation}
   \frac{\widetilde{d} s_{\rm tot}}{d t}   = \frac{\Phi_\ell}{T}  - \frac{q_j}{T^2} \frac{\partial T}{\partial x_j} -  \frac{\partial}{\partial x_j} \left( \frac{q_j}{T} \right),
 \label{entropytot}
\end{equation}
where we have rewritten $T^{-1}\nabla \cdot {\bf q} = T^{-2} {\bf q} \cdot \nabla T + \nabla \cdot ({\bf q}/T)$. The first two terms on the RHS of Eq. \ref{entropytot}  represent the   entropy generation terms (strictly positive in equilibrium thermodynamics due to the second law), while the last one represents spatial diffusion of entropy thus not associated with net generation. 

To complete the thermodynamic analogy, we identify the temperature to be the (internal) kinetic energy of the small-scale turbulence, i.e., we set  $T = k_\ell$ (in other words, we select a ``Boltzmann constant'' of unity thus choosing units of temperature equal to those of turbulent kinetic energy per unit mass). Examining Eq. \ref{entropytot} it is then quite clear that the quantity
\begin{equation}
   \widehat{\Psi}_\ell  = \frac{\Phi_\ell}{k_\ell}  - \frac{q_j}{k_\ell^2} \frac{\partial k_\ell}{\partial x_j}
 \label{entropygen1}
\end{equation}
represents the total entropy generation rate for the system formed by the smaller-scale eddies inside any particular sphere of diameter $\ell$.  In this paper we do not focus on the entropy generation due to spatial gradients in small-scale kinetic energy (the second term in Eq. \ref{entropygen1}) and focus solely on the part due to the cascade of kinetic energy in scale space, 
\begin{equation}
   \Psi_\ell  = \frac{\Phi_\ell}{k_\ell}.  
 \label{entropygen2}
\end{equation}

  The structure function formalism leading to $\Phi_\ell$ as the quantity describing the rate of local energy cascade at scale $\ell$ is not the only formalism that can be used to quantify cascade rate in turbulence. Another approach is widely used in
the context of Large Eddy Simulations (LES), where an equation similar to Eq. \ref{KHMH_local2} can be obtained  using filtering \citep{piomelli1991subgrid,germano1992turbulence,meneveau2000scale}. It is a transport equation for the trace of the  subgrid-scale or subfilter scale stress tensor $\tau_{ij}  = \widetilde{u_i u}_j-\tilde{u}_i\tilde{u}_i$ (the {\it tilde} $\widetilde{...}$ represents spatial filtering at scale $\ell$) i.e., a transport equation for $k^{\rm sgs}_\ell = \frac{1}{2} \tau_{ii}$. In this equation the term $\Pi_\ell = - \tau_{ij} \tilde{S}_{ij}$ 
appears ($\tilde{S}_{ij}$ is the filtered strain-rate tensor), and $\Pi_\ell$ plays a role similar to the role of $\Phi_\ell$ for the velocity increment (or structure function) formalism (see \cite{yao2023comparing} for a comparative study of both). Consequently, we can define another entropy generation rate associated with subgrid or subfilter-scale motions according to $\Psi^{\rm sgs}_\ell  = {\Pi_\ell}/{k^{\rm sgs}_\ell}$.

In any case, the system consisting of the small-scale eddies inside the sphere of diameter $\ell$ cannot be considered to be in near statistical equilibrium and thus $\Phi_\ell$ or $\Pi_\ell$ (and $\Psi_\ell$ or $\Psi_\ell^{\rm sgs}$) can in principle be both positive and negative. In particular, the literature on observations of negative subgrid-scale energy fluxes $\Pi_\ell$ is extensive  \citep{piomelli1991subgrid,borue1998local,meneveau2000scale,
van2002effects,vela2022subgrid}. The lack of equilibrium conditions in turbulence is related to the fact that there is no wide time-scale separation between the eddies smaller than $\ell$ and those at or larger than $\ell$. It is also related to the fact that   the number of entities, eddies, or ``particles'', at scales smaller than $\ell$ that are dynamically interacting with those at scales larger than $\ell$ is not large as it is in molecular systems. Therefore, local violations of analogues of the second law are to be expected and relevant principles from non-equilibrium thermodynamics must be invoked instead.
   We regard the evolution of small-scale eddies at scales below $\ell$, but significantly larger than the Kolmogorov scale, as being governed by inviscid, reversible dynamics. These eddying degrees of freedom would be the analogue of the reversible dynamics of molecular degrees of freedom at the microscopic level. The reversible microscopic dynamics of such molecules give rise to positive definite dissipation rate $\epsilon$ and phase-space volume contraction when motions are coarse-grained at continuum description scales. For the turbulence case, we posit that phase-space contraction and entropy generation occurs at the level of coarser-grained dynamics, when attempting to describe the system using effective variables at scales at, or larger than, $\ell$. The reversible inviscid eddying motions at scales smaller than $\ell$ give rise to $\Phi_\ell$ in analogy to how reversible microscopic molecular dynamics give rise to $\epsilon$. However, because of the lack of scale separation between the small-scale eddies and $\ell$,  $\Phi_\ell$ and phase-space volume change for variables at that level of description can be either positive or negative. 

  It should be kept in mind that defining entropy for non-equilibrium systems is in general not a settled issue, even for fields other than fluid turbulence.  For the purpose of exploring the consequences of a relatively simple option, we follow the definition used in equilibrium systems as in Eq. \ref{eq:TDS}. Clearly, since $\Phi_\ell$ and $\Pi_\ell$ can be negative, so will the entropy generation rates, and second law violations will be possible using the currently proposed definition of entropy.

\section{Fluctuation Theorem in non-equilibrium thermodynamics}
\begin{figure}
 \centering
  \includegraphics[scale=0.18]{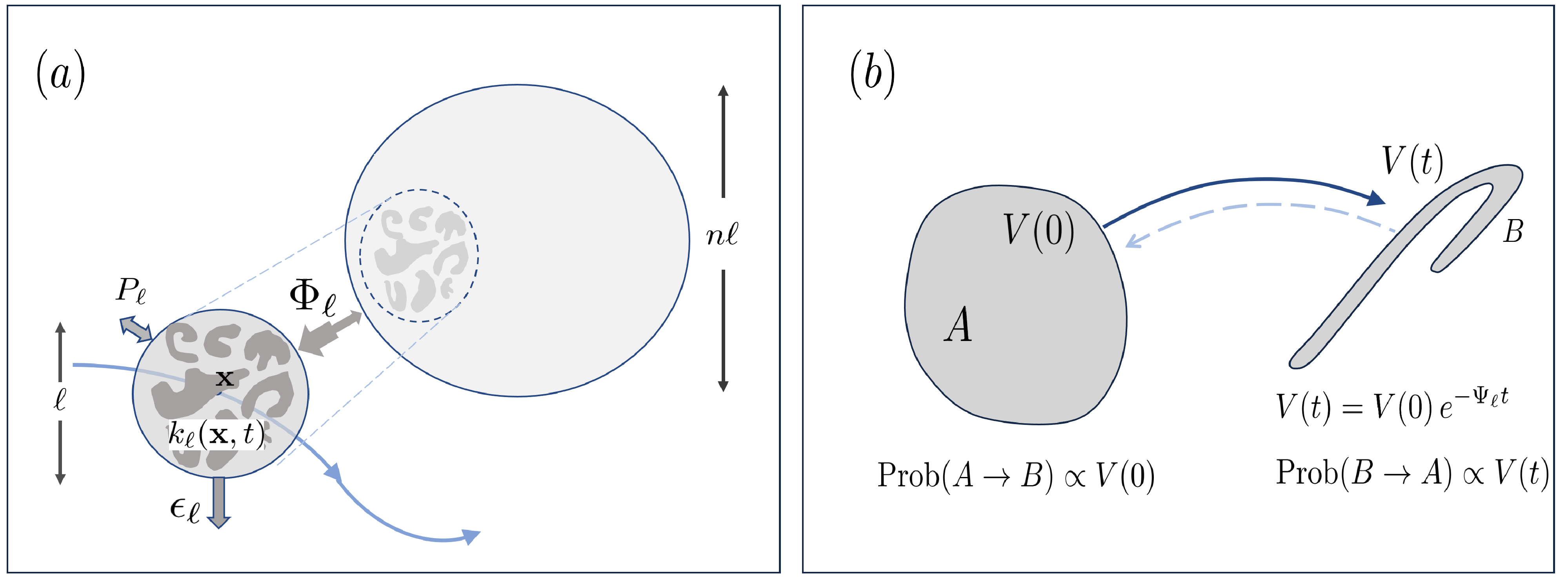}
    \caption{(a) Sketch in physical space illustrating  eddies at scales $\ell$ and smaller being transported by the larger-scale flow and exchanging energy locally at a rate $\Phi_\ell$ with eddies of larger size ($n \ell$), and being affected by pressure work $P_\ell$. There is dumping of energy with a ``heat reservoir'' at a rate $\epsilon_\ell$. (b) Sketch in phase space representing the (``microscopically'' reversible) dynamics of a set ($A$) of possible states of the system  that are characterized by phase-space contraction rate $\Psi_\ell$, that start at $t=0$ and evolve to states $B$ at time $t$. The ``microscopic'' degrees of freedom here are the eddies of scale smaller that $\ell$ and in the inertial range their dynamics are reversible.}
    \label{fig:sketchab}
\end{figure}

A well-known and testable result from non-equilibrium thermodynamics is the Fluctuation Relation, FR \citep{evans1993probability,gallavotti1995dynamical,searles2000generalized,marconi2008fluctuation,seifert2012stochastic}. Very loosely speaking, for systems in which the microscopic dynamics are reversible (as they can be argued to be in the case of small-scale eddies in the inertial range obeying nearly inviscid dynamics), the ratio of probability densities of observing a ``forward positive dissipative'' event and the same ``negative dissipation reverse'' event can be related to the contraction rate in the appropriate phase-space. The sketch in Fig. \ref{fig:sketchab}(b) illustrates the evolution of a ``blob'' of states of the system (set of states ``A'' occupying volume $V(0)$ in phase space) at time $t=0$. These states evolve and after some time $t$ the corresponding phase-space volume has changed to $V(t)$ and the set of states now occupies set $B$.

On average due to positive mean entropy generation and associated contraction of phase-space volume, $V(t)<V(0)$, but for certain configurations the reverse may be true. The probability of observing one of the states in set $A$ can be taken to be proportional to the phase-space volume of set $A$. Thus, the probability of being in set $A$ (and therefore ending up in $B$ after a time $t$) is proportional to $V(0)$, i.e., $P(A\to B) \sim V(0)$.  Phase-space contraction rates involve exponential rates of volume change depending on the finite-time Lyapunov exponents. Since the phase-space contraction rate in dynamical systems is proportional to the local rate of entropy generation ($\Psi_\ell$ in our case), one expects $V(t) = V(0) \exp(-\Psi_\ell t)$,
assuming that the initial set A was chosen specifically to consist only of sets of states characterized by $\Psi_\ell$ between times $t=0$ and $t$. Crucially, since the dynamics are reversible, if one were to run time backwards and start with the states at $B$, one would end up at $A$. Also, $P(B\to A) \sim V(t)$. The corresponding entropy production rate would have the opposite sign (as $\Psi_\ell$ is an odd function of velocities). Identifying $P(A\to B)$ with the probability density $P(\Psi_\ell)$ of observing a given value of entropy generation and 
$P(B\to A)$ with the probability density of observing the sign-reversed value, i.e. $P(-\Psi_\ell)$, leads to the FR relationship applied to the entropy generation rate defined for our turbulence system:
\begin{equation}
   \frac{P(\Psi_\ell)}{P(-\Psi_\ell)} = \frac{V(0)}{V(t)} = \exp\left( \Psi_\ell \, t \right),
 \label{FRrelation1}
\end{equation}
where time $t$ is understood as the time over which the entropy generation rate is computed if in addition one were to average over periods of time following the sphere of size $\ell$ in the flow. For now we shall not assume a specific value of $t$ and assume it is small but finite.  If Eq. \ref{FRrelation1} holds true in turbulent flows, a plot of $\log[P(\Psi_\ell)/P(-\Psi_\ell)]$ versus $\Psi_\ell$ should show linear behavior when plotted as function of $\Psi_\ell$.  

\section{Results from isotropic turbulence at $R_\lambda=1250$}

To evaluate the validity of the FR for isotropic turbulence we use data from a direct numerical simulation (DNS) of forced isotropic turbulence at a Taylor-scale Reynolds number of $R_\lambda = 1{,}250$. The simulations used 8,192$^3$ grid points \citep{yeung2012dissipation} and the data are available at the public Johns Hopkins Turbulence Database system (JHTDB). We perform the analysis at three length-scales in the inertial range, $\ell=30\eta, \,45\eta, \,60\eta$ where $\eta$ is the average Komogorov scale.   
To compute surface averages required to evaluate $\Phi_\ell$, we discretize the outer surface of diameter $\ell$ into 500 point pairs ($+$ and $-$ points) that are approximately uniformly distributed on the sphere. Velocities 
to evaluate $\delta u_i$ are obtained using the JHTDB webservices. $k_{\ell}$ is evaluated similarly by integrating over five concentric spheres. 
The accuracy of this method of integration has been tested by increasing the number of points used in the discretization.  To compute spherically volume filtered quantities such as $\tau_{ij}$ or $\tilde{S}_{ij}$, we fix the middle point coordinate ${\bf x}$ in the physical domain. For each center point, we download data in a cubic domain using the JHTDB's cutout service in a cube of size equal to $\ell$. The arrays are then multiplied by a spherical mask (filter) to evaluate local filtered velocities and velocity products. Gradients are evaluated using 4th-order centered finite differences. 
We compute the quantities $k_\ell$, $\Phi_\ell$ and $\Psi_\ell = \Phi_\ell/k_\ell$ at $2\times 10^6$ randomly chosen points in the domain. The probability density functions of $\Psi_\ell$ (and of $\Phi_\ell$ ) are then evaluated based on the entire sample of randomly chosen points.

\begin{figure}
 \centering
  \includegraphics[scale=0.35]{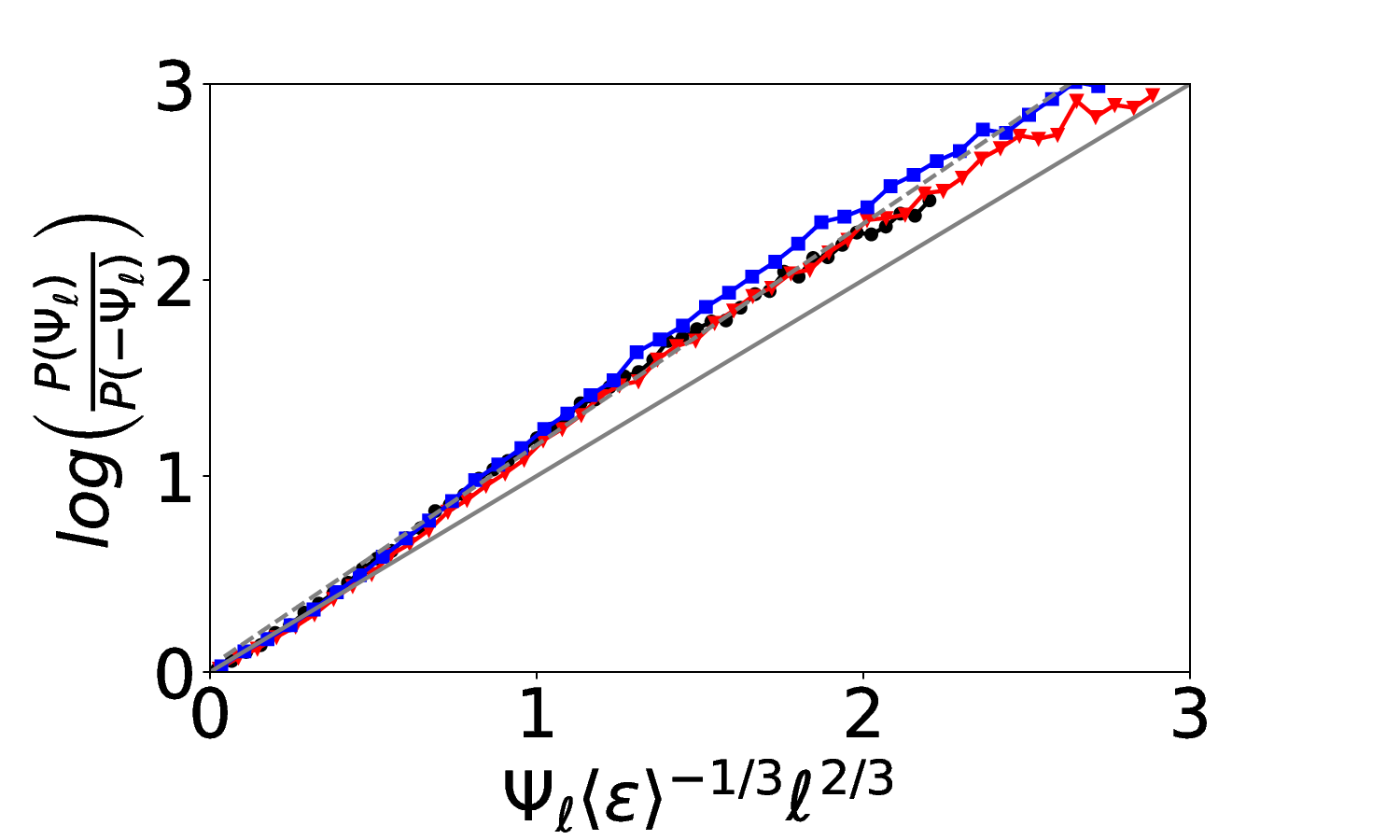}
    \caption{Fluctuation Relation test for isotropic turbulence at $R_\lambda=1250$: ratio of probability densities of positive and negative entropy generation rate scales exponentially with the entropy generation rate $\Psi_\ell$ at scale $\ell$. Results are shown for 3 different scales $\ell/\eta = 30$ (black circles), 45 (red triangles) and 60 (blue squares).  The gray dashed line has slope=1.13 obtained via linear fit while the solid gray line has slope = 1. In this and all other figures, natural logarithm is used.}
    \label{fig:FRresult}
\end{figure}

Figure \ref{fig:FRresult} shows the ratio of probability densities for positive and negative entropy production rates as function of the entropy production rate $\Psi_\ell$, in semi-logarithmic axes. Results are shown for three scales $\ell/\eta = 30, 45$ and 60. In good agreement with the prediction of the fluctuation relation, to a good approximation the results show linear behavior, over a significant range of $\Psi_\ell$ values. The units of $\Psi_\ell$ are inverse time-scale, so that they are here normalized by the inertial range scaling of this quantity, $\langle \epsilon \rangle^{1/3} \ell^{-2/3}$. 

The slope of the lines, when $\Psi_\ell$ is normalized by $\langle \epsilon \rangle^{1/3} \ell^{-2/3}$ is rather independent of $\ell/\eta$ and is quite close to unity. It suggests that the elapsed ``time'' is on the order of $t \sim \tau_\ell$, where $\tau_\ell= \langle \epsilon \rangle^{-1/3} \ell^{2/3}$, consistent with the notion of eddy turnover-time.  Figure \ref{fig:FRresult} represents the main finding of this study, providing strong support for the applicability of FR in the context of turbulence in the inertial range, provided the entropy generation rate is defined  based on the ratio of energy cascade rate and local ``temperature'' $k_{\ell}$. 

Furthermore, if one were to interpret the normalized entropy generation rate as an entropy change, i.e. $ \Delta s = \Psi_\ell \tau_\ell$,  one can also test the integral fluctuation relation \citep{marconi2008fluctuation,seifert2012stochastic,fuchs2020small} which states that $\langle \exp(-\Delta s)\rangle = 1$. Remarkably,  computing the average over all $N=2\times 10^6$ samples, we obtain $\langle \exp(-\Psi_\ell \langle \epsilon \rangle^{-1/3} \ell^{2/3})\rangle=0.99$, $1.03$ and $0.97$ at the three scales $\ell/\eta =$ 30, 45 and 60, respectively. Statistical convergence of our evaluation of $\langle e^{-\Delta s} \rangle$ is very good: For the case where $\ell/\eta =$ 30 and $N=0.5\times 10^6$ and $10^6$, the corresponding $\langle e^{-\Delta s} \rangle$ are 0.9856 and 0.9845, respectively. Similarly for $\ell/\eta =$ 45 and 60 the disparity is less than $1\%$

This confirmation of the validity of the integral fluctuation relation suggests that $\tau_\ell = \langle \epsilon\rangle^{-1/3} \ell^{2/3}$ is the natural timescale for the cascade process,  although $\tau_\ell$ corresponds to an average turn-over timescale (since it is based on the global mean dissipation instead of the local dissipation $\epsilon_\ell$). $\tau_\ell$ may therefore be interpreted as describing the level in the cascade process corresponding to scale $\ell$ (as envisioned in the approach by \cite{fuchs2020small}) rather than representing the actual elapsed time during an eddy turnover process, for which the local time based on $\epsilon_\ell$ could be more appropriate  (for analysis of conditional statistics based on $\epsilon_\ell$, see \cite{yao2023forward}).

\section{Discussion}
\begin{figure}
 \centering
  \includegraphics[scale=0.26]{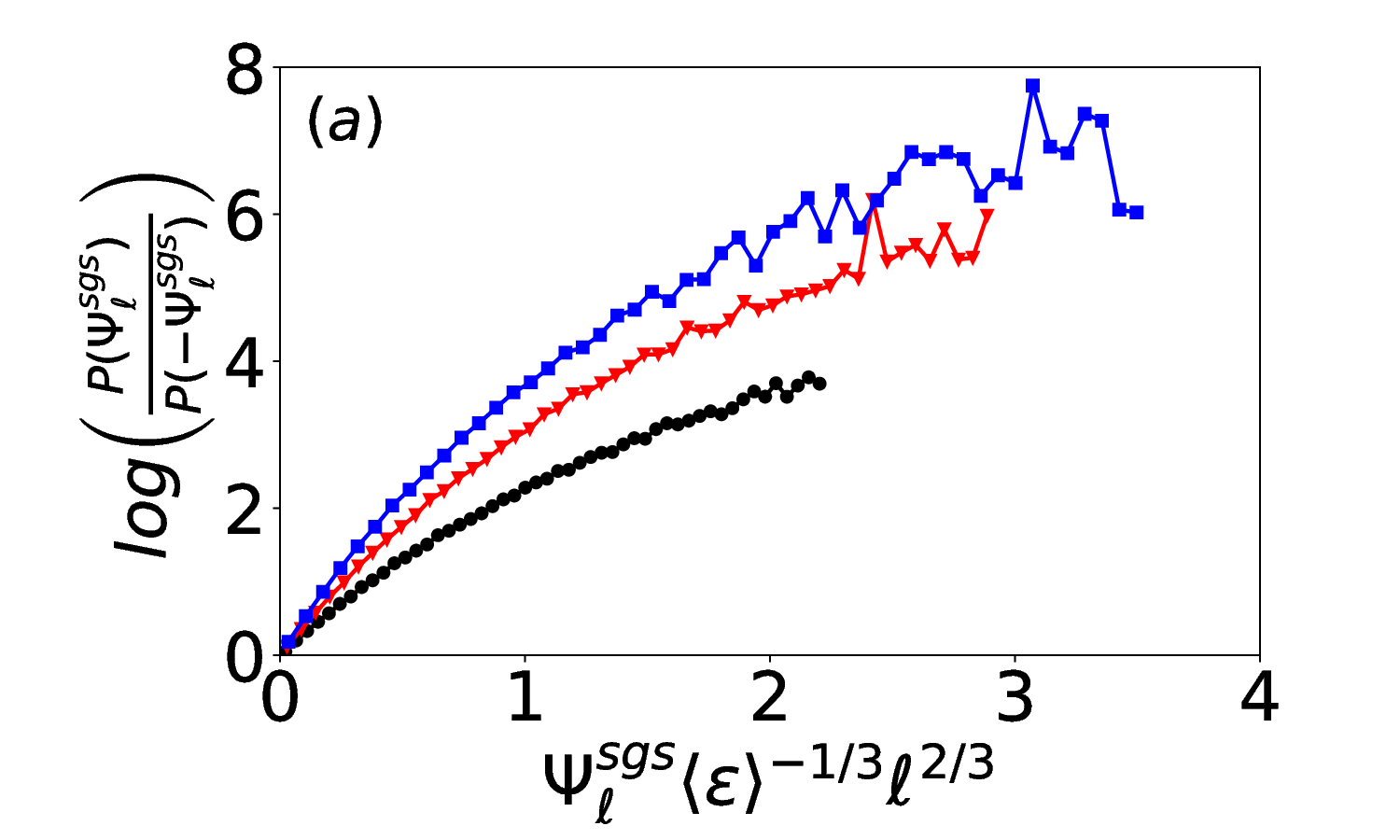}
  \includegraphics[scale=0.26]{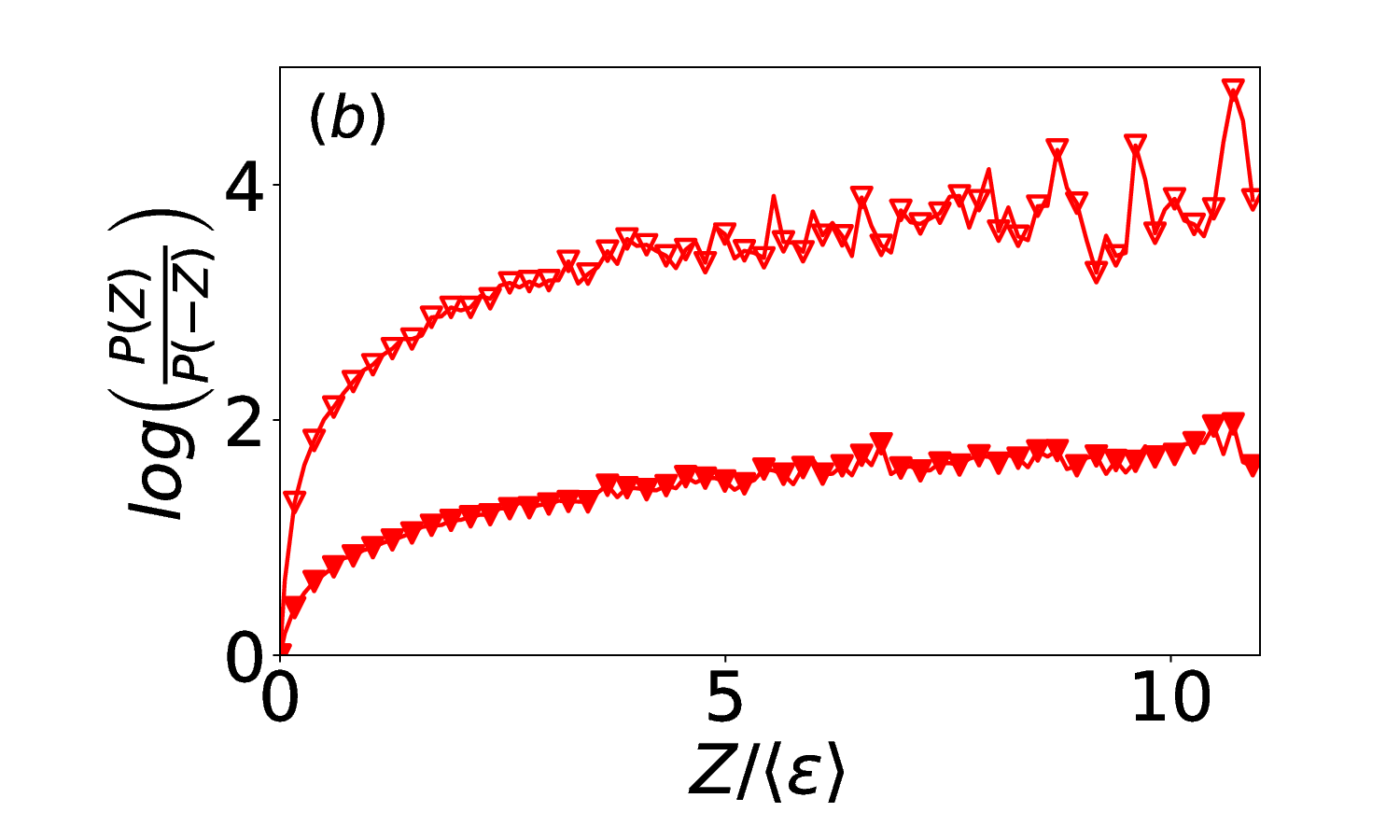}
    \caption{(a) Fluctuation Relation test for isotropic turbulence at $R_\lambda=1250$ applied to the entropy generation rate suggested by the LES filtering formalism $\Psi_\ell^{sgs}=\Pi_\ell/(\tau_{ii}/2)$ (normalized by the timescale $\tau_\ell = \langle \epsilon\rangle^{-1/3} \ell^{2/3}$) for 3 filtering scales $\ell/\eta$=30 (black circles), 45 (red triangles) and 60 (blue squares).  (b) Fluctuation Relation test applied to the cascade rates $Z = \Phi_\ell$ (solid red trangles) and $Z = \Pi_\ell$ (open red trangles) directly, without division by local kinetic energy (``temperature''). Results are shown for scale $\ell/\eta$ =  45  but results for other scales are similar.}
    \label{fig:FRresultPiothers}
\end{figure}

Here we explore some other plausible quantities and entropy definitions, and test to what degree FR can apply to them. First, we test applicability of the FR relation to the entropy production rate $\Psi_\ell^{sgs}$ as suggested in the filtering formalism from LES. Figure \ref{fig:FRresultPiothers}(a) shows that the corresponding FR does not exhibit linear behavior, i.e. the FR does not apply to the LES version of entropy generation rate $\Pi_\ell/\frac{1}{2}\tau_{ii}$ (at least not for the scales $\ell/\eta$ studied here).  Another variant is motivated by considering directly the cascade rates $\Phi_\ell$ and $\Pi_\ell$ rather than $\Psi_\ell$ or $\Psi_\ell^{sgs}$ as representative of the entropy production rate. We remark that the identification of $\Pi_\ell$ as ``entropy generation rate'' is commonplace in the literature, presumably because a constant reference (arbitrary) temperature is assumed. Figure \ref{fig:FRresultPiothers}(b) shows that such definitions also do not exhibit linear behavior and thus the cascade rates do not obey the FR relations.  Our results show that $\Phi_\ell$ must be divided by $k_\ell$ (temperature) to properly correspond to an entropy generation rate (units of 1/time) and only then they exhibit behavior consistent with FR  (Fig. \ref{fig:FRresult}).  

\begin{figure}
 \centering
  \includegraphics[scale=0.26]{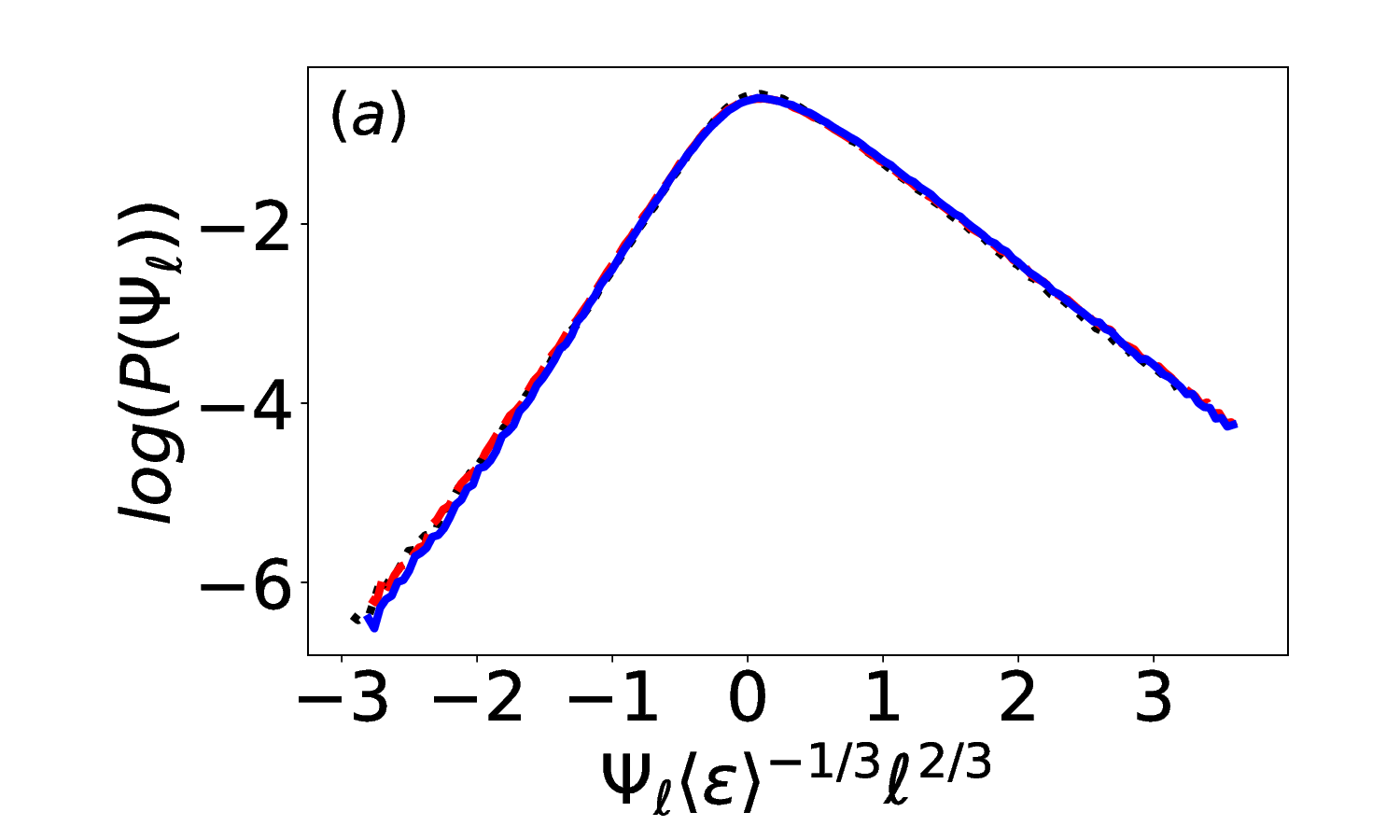}
  \includegraphics[scale=0.26]{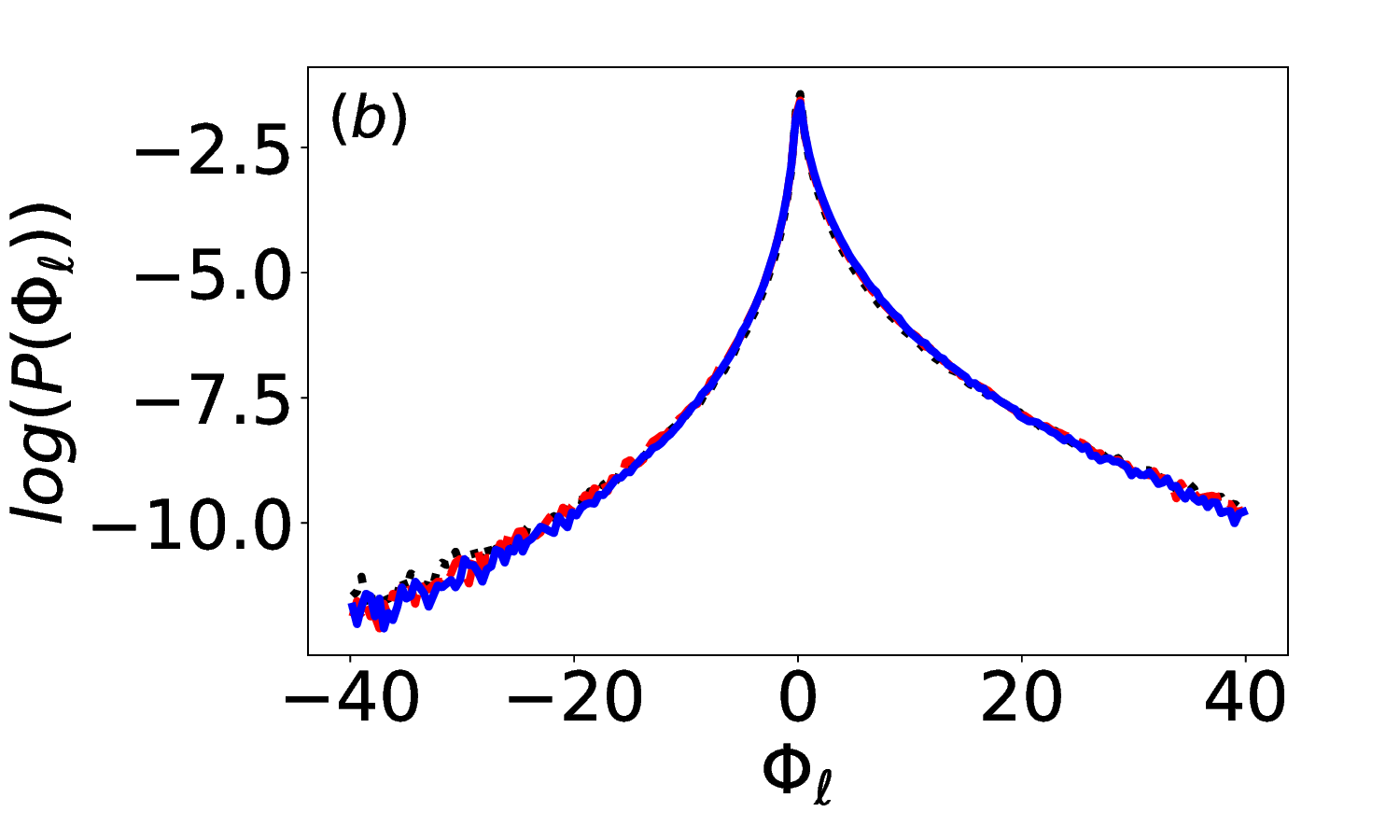}
    \caption{ (a) PDFs of entropy production rate $\Psi_\ell$. (b) PDFs of energy cascade rate $\Phi_\ell$ ($\Phi_\ell$ is shown in simulation units \citep{yeung2012dissipation}, for which $\langle \epsilon\rangle = 1.367$). Results are shown in semi-logarithmic axes, for 3 different scales $\ell/\eta = 30$ (black dotted line), 45 (red dashed line) and 60 (blue solid line).}
    \label{fig:PDFsPhiPsi}
\end{figure}

For more in-depth understanding of the observed trends, we show  PDFs of the 
entropy production rate $\Psi_\ell$ in Fig. \ref{fig:PDFsPhiPsi}(a) and the
energy cascade rate $\Phi_\ell$  in Fig. \ref{fig:PDFsPhiPsi}(b), all at the three scales $\ell$ (note that here we do not normalize $\Psi_\ell$ and $\Phi_\ell$ by their inertial range values).  
PDFs of energy cascade rate $\Pi_\ell$ have been shown in the literature on many occasions, especially for the filtering/LES formulations (see e.g., \cite{borue1998local,cerutti1998intermittency,tao2002statistical,cardesa2015temporal,vela2021entropy}). A detailed comparative study between statistics of $\Phi_\ell$ and $\Pi_\ell$ has been presented elsewhere \citep{yao2023comparing}. Here we note that the PDFs of $\Phi_\ell$ quantities have elongated highly non-Gaussian tails. Consistent with many prior observations \citep{borue1998local,cerutti1998intermittency,vela2021entropy} regarding the PDFs of $\Pi_\ell$, they have tails that are much wider (i.e., even more intermittent) than having exponential tails. However,  by considering the variable $\Psi_\ell$ (i.e., properly dividing by temperature), the tails of the PDF of $\Psi_\ell$ become visibly much closer to exponential. Extreme events of $\Psi_\ell$, once divided by the prevailing local kinetic energy, become less extreme. As can be seen in Fig. \ref{fig:PDFsPhiPsi}(a), the slopes of the exponential tails differ on the negative (steeper) and positive (flatter) sides. 

We note that if both sides of the PDF have an exponential tail (e.g., $P(\pm\Psi_\ell) \sim \exp(-\alpha_\pm |\Psi_\ell|)$ with $\alpha_+$ characterizing the positive $\Psi_\ell$ tail and $\alpha_-$ the negative one, the FR holds trivially and the slope of $\log(P(\Psi_\ell)/P(-\Psi_\ell))$ versus $\Psi_\ell$ is $\alpha_--\alpha_+$. For the case of normalization using $\tau_\ell$, we thus have $\alpha_--\alpha_+ \approx 1$, approximately independent of scale $\ell$ in the inertial range. For purely two-sided exponential PDFs with the two slopes $\alpha_-$ and $\alpha_+$, one can show that $\langle \exp(-\Psi_\ell \tau_\ell) \rangle = \alpha_- \alpha_+(\alpha_--1)^{-1}(\alpha_++1)^{-1}$ which equals unity if  $\alpha_--\alpha_+ = 1$, consistent with the integral fluctuation theorem.  These observations must be kept in mind when interpreting the results supporting the FR behavior seen in Fig. \ref{fig:FRresult}: On the one hand, as argued before, they could point to non-equilibrium thermodynamic behavior expected for systems far from equilibrium. Or, perhaps more mundanely, they could be a mere consequence of exponential tails in the PDFs of the ratio of energy transfer rate divided by local kinetic energy in turbulence. Perhaps both interpretations are non-trivially connected.

Returning to the proposed definition of entropy in Eq. \ref{KHMH_gibbs1} associated to the system of eddies in a sphere of diameter $\ell$, it is instructive to rewrite it in ``increment'' form (which again has to be interpreted in Lagrangian fashion) and it would read
\begin{equation}
    d s_\ell = \frac{1}{k_\ell} \left( d k_\ell + d w_\ell \right) = d\, \ln(k_\ell) + k_\ell^{-1} d w_\ell.
\end{equation}
Here $d w_\ell = (6/\ell) [(p^*/\rho) d \delta {\bf s} \cdot \hat {\bf r}]_{S_\ell}$ is the spherically averaged pressure work such that $\delta {\bf u} = \delta d{\bf s}/dt$. Whether this definition can somehow be related to the (log of) the number of possible states of the eddies smaller than $\ell$, or provide any additional predictive capabilities (besides the observed FR behavior), remains to be seen. 

As future extensions of the present study, it would be of interest to consider the effects of Reynolds number and scale $\ell$ approaching either the viscous or the integral scale of turbulence, to consider flows other than isotropic turbulence, and also to explore the contributions of ``spatial diffusive fluxes of kinetic energy'' due to spatial gradients of $k_\ell$ that according to Eq. \ref{entropytot} should also contribute, perhaps separately, to the total entropy generation rate. The small deviations from precisely linear behaviour seen in Fig. \ref{fig:FRresult} also deserve further more detailed study. Furthermore, the role of ``cascade time'' $t$ has to be clarified. An obvious possibility is to follow $\Psi_\ell$ in a Lagrangian frame \citep{meneveau1994lagrangian,wan2010dissipation} and perform additional finite-time averaging.  
\vspace*{-12pt}
\section*{Acknowledgements}
We thank  G. Eyink for fruitful comments and the JHTDB/IDIES staff for their assistance with the database and its maintenance. This work is supported by NSF (Grant \# CSSI-2103874). 

\vspace*{-15pt}
\section*{Declaration of interests}
The authors report no conflict of interest.

\vspace*{-12pt}
\bibliographystyle{jfm}
 
\bibliography{Ref}

\end{document}